\documentclass[12pt]{article}

\usepackage{amssymb,amsthm,amsmath,slashed}
\usepackage[T1]{fontenc}
\usepackage[cp1250]{inputenc}
\usepackage{setspace}

\newcommand{\as}{\mathrm{as}}
\newcommand{\inc}{\mathrm{in}}
\newcommand{\out}{\mathrm{out}}
\newcommand{\ret}{\mathrm{ret}}

\newcommand{\al}{\alpha}
\newcommand{\ep}{\epsilon}
\newcommand{\ga}{\gamma}
\newcommand{\la}{\lambda}
\newcommand{\mR}{\mathbb{R}}
\newcommand{\dsp}{\displaystyle}
\newcommand{\txt}{\textstyle}
\newcommand{\vep}{\varepsilon}

\newcommand{\w}{\omega}

\newcommand{\ov}{\overline}

\newcommand{\p}{\partial}
\newcommand{\dV}{\dot{V}}
\newcommand{\tdV}{\widetilde{\dot{V}}}
\newcommand{\ino}{\!\!\begin{smallmatrix}\inc\\\out\end{smallmatrix}\!\!}
\newcommand{\rea}{\!\!\begin{smallmatrix}\ret\\\adv\end{smallmatrix}\!\!}

\DeclareMathOperator{\Ip}{Im} \DeclareMathOperator{\Rp}{Re}

\title{Infrared limit in external field scattering}
\author{Andrzej Herdegen\thanks{e-mail: herdegen@th.if.uj.edu.pl}\\
{\it Institute of Physics, Jagiellonian University,}\\
{\it Reymonta 4, 30-059 Kraków, Poland}}
\date{}

\begin{document}

\maketitle

\begin{abstract}
Scattering of electrons/positrons by external classical electromagnetic wave packet is considered in infrared limit. In this limit the scattering operator exists and produces physical effects, although the scattering cross-section is trivial.

\vspace{1ex}
\noindent
keywords: QED; infrared problem; charged states
\end{abstract}

\section{Introduction}\label{int}

Perturbative quantum electrodynamics is an unquestionable pragmatical success. However, there are more fundamental structural questions in QED which are not fully understood. Among them is the infrared structure, which includes group of problems related to the characterization of physical charged states and particles, long-range structure, Gauss law etc.

The usual way of handling these questions in perturbative approach, whether purely pragmatical or more fundamentally oriented, is based on elimination, in the first step, of the problems by a distortion of the theory (locality by Gupta-Bleuler formalism, combined with finite photon mass, infrared cut-off or spacetime smearing function for interaction Lagrangian). In the pragmatical approach one then  restricts attention to scattering cross-sections, which may be handled by procedures described in standard textbooks. In more fundamental approaches one strives to remove the distortions and obtain truly physical states, including scattering theory (the most ambitious attempt is by Steinmann \cite{ste}). However, it is not clear, whether in the process of IR regularization some information is not lost, which substantiates alternative attempts.

Essentially, alternatives strive to include the long-range structure of electrodynamics from start. Here is the place of standard Coulomb gauge QED, which, however, has not been proven renormalizable. Here, also, are other less orthodox attempts as those by Gervais and Zwanziger \cite{zwa} or Staruszkiewicz \cite{sta} (see also \cite{her05}). In similar spirit the present author has proposed an algebraic model for asymptotic fields in QED \cite{her98} which respects Gauss law,  which was found to have interesting features \cite{her08}, and may be potentially considered as a starting point for a perturbation expansion.

The present note does not refer directly to this aim, but is intended as a step towards this long-term goal. We consider scattering of the  quantum Dirac field in the external electromagnetic field. This is a textbook problem, but we are here interested in an application not normally intended: the external electromagnetic field will represent a free wave packet typical for scattering situations. By taking an infrared limit (to be specified precisely below) we want to extract long-range effects in this setting. Our results will confirm on this more elevated level the phase transformation discovered for a wave-function in semi-classical approximation by Staruszkiewicz \cite{sta81}. Our setting may also throw light on an extension to full QED.

\section{Preliminaries}

A convenient representation of a solution of the wave equation has the form~\cite{her95}
\begin{equation}\label{lor}
 A(x)=-\frac{1}{2\pi}\int \dV(x\cdot l,l)\,d^2\!l\,.
\end{equation}
Here:
\begin{list}{\labelitemi}{\leftmargin=1.5em}
\item $l$ is a vector on the future lightcone;
\item $V(s,l)$ is a (sufficiently smooth) real function of a real variable $s$ and of $l$, $\dV(s,l)\equiv\tfrac{\partial}{\partial s}V(s,l)$, and $V$ is homogeneous of degree $-1$ in its arguments: for $\lambda>0$
\begin{equation}
 V(\lambda s,\lambda l)=\lambda^{-1}V(s,l),
\end{equation}
so that $\dV$ is homogeneous of degree $-2$;
\item $d^2\!l$ is the conformally invariant measure over null directions on the cone, which is applicable to integrands  homogeneous of degree $-2$. In particular, if $t$ is any timelike, future-pointing, unit vector and $l$'s are scaled to $t\cdot l=1$, then $d^2\!l$ is the standard measure on the sphere $l^2=0$, $t\cdot l=1$.
\end{list}
If $A(x)$ is a Lorentz vector potential of a free electromagnetic field then $V$ is a vector function and the Lorentz condition $\partial\cdot A(x)=0$ is guaranteed by the constraint \mbox{$l\cdot V(s,l)=0$}. In order to obtain a class of fields typical for scattering situations one has to restrict the class of $V$ functions. Let
\begin{equation}
 \tdV(\w,l)=\frac{1}{2\pi}\int\dV(s,l)e^{i\w s}ds\,.
\end{equation}
If there exists the limit $\dsp\tdV(0^+,l)=\lim_{\w\searrow0}\tdV(\w,l)$
then for all vectors $x$ and each spacelike $y$ one has \cite{her05}
\begin{multline}
 A_\as(y)=
 \lim_{R\to\infty}RA(x+Ry)\\=-\int\Rp[\tdV(0^+,l)]\,
 \delta(y\cdot l)\,d^2\!l-
 \frac{1}{\pi}\int\frac{\Ip[\tdV(0^+,l)]}{y\cdot l}\,d^2\!l\,,
\end{multline}
thus the potential has a tail of Coulomb-like decrease, whose shape is independent of the choice of the origin. However, not all fields of this decrease naturally appear in scattering situations: one shows that potentials of radiation fields produced by currents due to particles or fields which are free at asymptotic times (in and out) are characterized by long-range tails which are even functions of $y$. This leads to the restriction $\Ip[\tdV(0^+,l)]=0$. As from reality of $V(s,l)$ one has at the same time $\ov{\tdV(\w,l)}=\tdV(-\w,l)$, this condition is equivalent to the continuity of $\tdV(\w,l)$ at $\w=0$. In fact, a natural and  sufficient for the scattering context is the following more specific assumption: $\dV(s,l)$ is a smooth function which decays for $|s|\to\infty$ at least as $|s|^{-1-\ep}$ for some $\ep>0$. Then $V(s,l)$ has well-defined limits for $s\to\pm\infty$, and can be chosen so that
\begin{equation}
 \lim_{s\to\pm\infty}V(s,l)\equiv V(\pm\infty,l)=\pm\tfrac{1}{2}\Delta V(l)\,,\quad
 \Delta V(l)=2\pi\tdV(0,l)\,.
\end{equation}
Potentials characterized by such $V$'s have the null asymptotes given by
\begin{equation}
 \lim_{R\to\infty}RA(x\pm Rl)
 =\pm\big[\,V(x\cdot l,l)-V(\pm\infty,l)\,\big]\,.
\end{equation}

Radiation fields produced by scattered fields or particles are still more special than the class obtained above: their long-range tails are of electric type, which is equivalent to $y\wedge\p\wedge A_\as(y)=0$. One shows \cite{her98} that this is satisfied iff $l\wedge\p\wedge\Delta V(l)=0$, and then
\begin{equation}\label{Phi}
 l\wedge\Delta V(l)=-l\wedge\p\Phi(l)\,,\qquad
 \Phi(l)=-\frac{1}{4\pi}\int\frac{l\cdot\Delta V(l')}{l\cdot l'}\,d^2\!l'\,;
\end{equation}
note that $\Phi(l)$ is homogeneous of degree $0$, and under the gauge transformation $V(s,l)\to V(s,l)+l\al(s,l)$ changes merely by an additive constant.

The above characterization of the free potentials is related to the more usual Fourier representation
\begin{equation}
 A(x)=\frac{1}{\pi}\int a(k)\delta(k^2)\,\mathrm{sgn}(k^0)e^{-ix\cdot k}\,d^4\!k
\end{equation}
by
\begin{equation}
 a(\w l)=-\frac{\tdV(\w,l)}{\w}\,.
\end{equation}
Thus $a(k)$ may be singular of order $1/k^0$ at the cone vertex, but $\w a(\w l)$ is continuous across the vertex in $\w$. This restriction is responsible for the admissibility of what follows.

Note that $\tdV(\w,l)$ is continuous and fast decreasing in $\w$. We assume for simplicity that $\dV(s,l)$ is fast decreasing in $s$, and then $\tdV(\w,l)$ is smooth -- the infrared behavior of the field is not restricted by this assumption.

\section{Wave operators}\label{class}

Consider the Dirac equation for classical Dirac field in external potential
\begin{equation}\label{dir}
 \big[\,\ga\cdot(i\p-eA(x))-m\,\big]\psi(x)=0\,.
\end{equation}
It has been rigorously shown in \cite{her95} that this equation, when restricted to the inside of the future or past lightcone, has well-defined asymptotic free fields even for a class of long-range potentials, which includes potentials defined in the previous section. This allows one to write equivalent forms of this equation in these two respective regions
\begin{equation}
 \psi(x)=\psi_{\ino}(x)-e\int S_{\rea}(x-y)\ga\cdot A(y)\psi(y)\,d^4\!y\,,
\end{equation}
and then $\psi$ tends to $\psi_{\ino}$ in appropriate sense over the hyperboloids $x^2=\la^2$. More precisely, if we write free fields as
\begin{equation}\label{psifree}
 \psi_{\ino}(x)=\left(\frac{m}{2\pi}\right)^{3/2}
 \int e^{\txt -imx\cdot v\, \ga\cdot v}\ga\cdot v\,f_{\ino}(v)\,d\mu(v)\,,
\end{equation}
where $d\mu(v)$ is the invariant measure over the future unit hyperboloid, then
\begin{multline}\label{asym}
 \lim_{\la\to\infty}\mp i\la^{3/2}
 e^{\txt \mp i(m\la+\pi/4)\ga\cdot v}\psi(\mp\la v)\\
 = \lim_{\la\to\infty}\mp i\la^{3/2}
 e^{\txt \mp i(m\la+\pi/4)\ga\cdot v}\psi_{\ino}(\mp\la v)
 =f_{\ino}(v)\,,
\end{multline}
the limit in the Hilbert-norm sense on the hyperboloid, with scalar product given by $(f,g)=\int\ov{f}\ga\cdot vg(v)d\mu(v)$ (here and below the bar accent denotes  the usual Dirac conjugation).

We now want to relax somewhat the demands of rigor, assume the validity of the integral forms of equation for the whole space and treat them perturbatively. If one uses the representation \eqref{lor} and the Fourier representation
\begin{equation}
 S_{\rea}(x)=-\frac{1}{(2\pi)^4}\int M_{\rea}(p)e^{-ip\cdot x}d^4\!p\,,\quad
 M_{\rea}(p)=\frac{\ga\cdot p+m}{p^2-m^2\pm i0p^0}\,,
\end{equation}
then one finds
\begin{equation}\label{W}
\psi(x)=\int W_{\ino}(x,v) f_{\ino}(v)\,d\mu(v)\,,\quad
W_{\ino}(x,v)=\sum_{n=0}^\infty W_{\ino}^{(n)}(x,v)\,,
\end{equation}
\begin{multline}\label{Wn}
W_{\ino}^{(n)}(x,v)
=\left(\frac{-e}{2\pi}\right)^n\left(\frac{m}{2\pi}\right)^{3/2}
\sum_{\vep=\pm}\int e^{\txt -ix\cdot(\vep mv+\w_1l_1+\ldots+\w_nl_n)}\times\\
M_{\rea}(\vep mv+\w_1l_1+\ldots+\w_nl_n)\ga\cdot\tdV(\w_1,l_1)
M_{\rea}(\vep mv+\w_2l_2+\ldots+\w_nl_n)\ga\cdot\tdV(\w_2,l_2)\times\\
\times\ldots M_{\rea}(\vep mv+\w_nl_n)\ga\cdot\tdV(\w_n,l_n)\vep P_\vep(v)\,d^n\!\w\, d^{2n}\!l\,,
\end{multline}
where $P_\pm(v)=\tfrac{1}{2}(1\pm\ga\cdot v)$ project onto positive/negative frequency parts.
For fixed $x^0$ the operator $W_{\ino}(x,v)$ is a wave operator expressing field at a~given time in terms of incoming/outoing field characteristic. The one-particle scattering operator is thus given by
\begin{equation}\label{one-part}
 f_\out(v)=\int S_1(v,v')f_\inc(v')\,d\mu(v')\,,\quad
 S_1(v,v')=W_\out^*(x^0,.;v)W_\inc(x^0,.;v')\,.
\end{equation}

Formulae \eqref{W} -- \eqref{one-part} give a formal expansion of the one-particle wave- and scattering operators, but the language used above does not seem best-suited to go outside the perturbation theory. On the other hand, by formulating the Dirac equation inside the light-cones as an evolution over hyperboloids \mbox{$x\cdot x=\tau^2$} it was shown in \cite{her95}, as mentioned above, that $W_{\inc}(x,v)$ (resp. $W_{\out}(x,v)$) are defined non-perturbatively for $x$ in the past- (resp. future-) lightcone. It would be interesting to close the gap between them by considering the evolution on a foliation of Cauchy surfaces parametrized by $\tau$ and tending to $x\cdot x=\tau^2$, $x^0\gtrless0$, for $\tau\to\pm\infty$ respectively. Preliminary work was done in this direction in unpublished thesis \cite{mar}. It could then also be investigated whether the mixing parts of the one-particle scattering operator (those interpolating between positive and negative energy solutions) have finite Hilbert-Schmidt norm; this is the well-known necessary and sufficient condition for the scattering to be transferable to the quantum Dirac field (see e.g. \cite{sch}).

With regard to the last mentioned question it could seem that the answer is bound to be negative: typical sufficient conditions, as those formulated in theorem 5.1 of Scharf's book \cite{sch}, are evidently violated by free wave packets, as the 3-space transform of the potential oscillates harmonically in time. Also, for long-range fields, as radiation fields produced in scattering situations, the transform of the potential is singular of second power order in $\vec{p}=0$, which is another obstacle. However, the theorem is not `if and only if' one, and both difficulties seem to have chances not to play role in the above formulation.

However, this being a speculation at the present stage, we want to proceed differently to extract infrared-limit effects. We first perform an infrared limit (defined below) at the classical field level, and only then proceed to quantize the field. As the quantum scattering operator for classical external background problem, if it exists, is anyway a simple transcription of the classical one-particle operator, this procedure has good chances to give a realistic result.

\section{Infrared limit}

Let us now in the above scheme replace the potential $A(x)$ by a scaled potential
\begin{equation}\label{scaling}
 A_\la(x)=\la^{-1}A(\la^{-1}x)\,,
\end{equation}
with the intention of taking large $\la$ limit. The effect of this scaling is that while the long-range tails of the potential and of the field strength remain  unaffected for all $\la$ (as they decay in spacelike directions as $x^{-1}$ and $x^{-2}$ respectively), the potential and the corresponding field strength tend uniformly to zero for $\la\to\infty$ (as the potential is smooth and bounded). Also, it is easily seen that the energy content of the scaled field is $\mathcal{E}_\la=\la^{-1}\mathcal{E}$ ($\mathcal{E}$~before the scaling), so it vanishes in the limit.

In terms of $V$'s the scaling has the form
\begin{equation}
 V_\la(s,l)=V(\la^{-1}s,l)\,,\quad \tdV_\la(\w,l)=\tdV(\la\w,l)\,.
\end{equation}
As the field vanishes uniformly in the limit it could seem that the equation \eqref{dir} becomes free and the scattering matrix trivializes. While the first of these statements is true (free field equation), the second is not (trivial scattering), and the error in this conclusion comes from non-interchangeability of the limits. We are interested in physics for large, but finite $\la$, thus the limit $\la\to\infty$ must only be taken on the level of physical quantities of interest. We want to find the limits of wave operators, so we replace in \eqref{Wn} (order by order) $\tdV\to\tdV_\la$. If this is followed by a change of variables $\la\w_i\to\w_i$ one finds that the scaled version of \eqref{Wn} is obtained by replacing $\w_i\to\la^{-1}\w_i$ in the exponent and in each $M_{\rea}$, and multiplying each $M_{\rea}$ by $\la^{-1}$. But if $\xi_k=\w_kl_k+\ldots+\w_nl_n$, one finds a well-defined distributional limit
\begin{equation}
 \la^{-1}M_{\rea}(\vep mv+\la^{-1}\xi_k)\to
 \frac{\vep P_\vep(v)}{v\cdot\xi_k\pm i0}\qquad \text{for}\quad \la\to\infty\,.
\end{equation}
Using this one finds that \eqref{Wn} takes in the limit the form
\begin{equation}
 W_\infty^{(n)}{}_{\ino}(x,v)=\left(\frac{m}{2\pi}\right)^{3/2}
 e^{\txt -imx\cdot v\ga\cdot v}\ga\cdot v \,R^{(n)}_{\ino}(v)\,,
\end{equation}
where
\begin{multline}\label{Rn}
 R^{(n)}_{\ino}(v)=\left(\frac{-e}{2\pi}\right)^n\int \frac{v\cdot\tdV(\w_1,l_1)}{v\cdot(\w_1l_1+\ldots+\w_nl_n)\pm i0}\,
 \times\\
 \frac{v\cdot\tdV(\w_2,l_2)}{v\cdot(\w_2l_2+\ldots+\w_nl_n)\pm i0}\,
 \ldots\ldots\,\frac{v\cdot\tdV(\w_n,l_n)}{v\cdot\w_nl_n\pm i0}\,
 d^n\!\w\,d^{2n}\!l\,.
\end{multline}
This means that $\psi$ becomes in this limit a free field
\begin{equation}\label{psilim}
 \psi(x)=\left(\frac{m}{2\pi}\right)^{3/2}
 \int e^{\txt -imx\cdot v\, \ga\cdot v}\ga\cdot v\,f(v)\,d\mu(v)\,,
\end{equation}
with
\begin{equation}\label{flim}
 f(v)=R_{\ino}(v)f_{\ino}(v)\,,
\end{equation}
where $\dsp R_{\ino}(v)=\sum_{n=0}^\infty R^{(n)}_{\ino}(v)$.
Going in \eqref{Rn} to the inverse Fourier transform and denoting
\begin{equation}
 \eta(\tau)=-\frac{1}{2\pi}\int\frac{v\cdot V(\tau v\cdot l,l)}{v\cdot l}\,d^2\!l
\end{equation}
we obtain
\begin{equation}
 R_{\ino}(v)=\sum_{n=0}^\infty(\mp ie)^n\int\dot{\eta}(\tau_1)\ldots
 \dot{\eta}(\tau_n)\theta(\mp\tau_1)\theta(\pm(\tau_1-\tau_2))\ldots
 \theta(\pm(\tau_{n-1}-\tau_n))\,d^n\!\tau\,.
\end{equation}
One recognizes in this formula the chronological/antichronological exponent of the integral
 $\dsp\mp ie\int_{\mR_\mp}\dot{\eta}(\tau)\,d\tau$. But as $\dot{\eta}$ is a numeric function, this reduces to ordinary exponent, and one finds
 \begin{equation}
  R_{\ino}(v)=\exp\left[\,
  \frac{ie}{2\pi}\int
  \frac{v\cdot[V(0,l)-V(\mp\infty,l)]}{v\cdot l}\,d^2\!l
  \,\right]\,.
\end{equation}
The one-particle scattering operator $S_1$ in infrared limit is thus the multiplication by a phase:
\begin{equation}\label{scalim}
 f_\out(v)=S_1(v)f_\inc(v)\,,\quad
 S_1(v)=\exp\left[\,
 \frac{ie}{2\pi}\int\frac{v\cdot\Delta V(l)}{v\cdot l}\,d^2\!l
 \,\right]\,.
\end{equation}

The formulas \eqref{psilim}, \eqref{flim} may seem paradoxical: if one calculates the asymptotic limits as in \eqref{asym} one gets $f$ instead of $f_{\ino}$. But, in fact, this only confirms that now this operation is not allowed: after taking the infrared limit ($\la\to\infty$) the asymptotic time limit need not reproduce the limit found before the scaling.

The phase transformation $S_1(v)$ has been discovered in the setting of semi-classical approximation for a wave-function by Staruszkiewicz \cite{sta81}. Note that for this limit of scattering operator \eqref{scalim} the choice of the origin in Minkowski space with respect to which the scaling \eqref{scaling} takes place is irrelevant; it is only the infrared tail which counts (which is invariant under translations). Finally, we note that with the use of \eqref{Phi} one can obtain \cite{her98} an equivalent form of the phase by the identity
\begin{equation}\label{VPhi}
 \int\frac{v\cdot\Delta V(l)}{v\cdot l}\,d^2\!l
 =-\int\frac{\Phi(l)}{(v\cdot l)^2}\,d^2\!l\,,
\end{equation}
which shows that the argument of the exponent in $S_1(v)$ is a bounded function of $v$.

\section{Quantum field}

The one-particle scattering operator \eqref{scalim} evidently satisfies the condition for quantum implementability, as the mixing parts are absent.
We can thus proceed to quantize the limit as described in closing remarks of Section \ref{class}.

Let the free asymptotic Dirac fields be represented in analogy to
\eqref{psifree} as
\begin{equation}
 \Psi_{\ino}(x)=\left(\frac{m}{2\pi}\right)^{3/2}
 \int e^{\txt -imx\cdot v\, \ga\cdot v}\ga\cdot v\,g_{\ino}(v)\,d\mu(v)\,,
\end{equation}
 with $g_{\ino}(v)$ combining the creation/annihilation operators for electron/posi\-tron,
\begin{equation}
 \big[g_{\ino}(v),g_{\ino}(v')\big]_+=0\,,\quad
 \big[g_{\ino}(v),\ov{g}_{\ino}(v')\big]_+=\ga\cdot v\,\delta_\mu(v,v')\,,
\end{equation}
where $\delta_\mu(v,v')$ is the delta function with respect to the measure $d\mu(v')$, and write
\begin{equation}
 g_{\ino}(f)=\int\ov{f}(v)\ga\cdot v\,g_{\ino}(v)\,d\mu(v)\,
\end{equation}
Then the $S$ operator is defined by
\begin{equation}
 g_\out(f)=S^*g_\inc(f)S\qquad \text{and here}\quad g_\out(f)=g_\inc(S^*_1f)\,.
\end{equation}
If one denotes $\rho_{\ino}(v)=\,:\!\ov{g}_{\ino}(v)\ga\cdot vg_{\ino}(v)\!:$ (normal ordering) and for a~measurable real function $\chi(v)$ writes $\rho_{\ino}(\chi)=\int\chi(v)\rho_{\ino}(v)\,d\mu(v)$ then
\begin{equation}
  \exp[i\rho_{\ino}(\chi)]\,g_{\ino}(f)\exp[-i\rho_{\ino}(\chi)]
  =g_{\ino}(e^{i\chi}f).
\end{equation}
Therefore
\begin{equation}\label{scop}
 S=\exp\left[\,
 \frac{ie}{2\pi}\int\frac{v\cdot\Delta V(l)}{v\cdot l}\,d^2\!l\,
 \rho_\inc(v)\,d\mu(v)\,\right]\,,
\end{equation}
where the operator in the exponent is bounded by the particle number operator (due to \eqref{VPhi}).
On the other hand for $A(x)$ given by \eqref{lor} and a current vector field $J(x)$ one has
\begin{equation}
 \int A(x)\cdot J(x)\,d^4\!x=-\frac{1}{2\pi}\int \dV(s,l)\cdot j(s,l)\,ds\,d^2\!l\,,
\end{equation}
where $j(s,l)=\int\delta(s-x\cdot l)J(x)\,d^4\!x$. For $J_\inc(x)=e:\!\ov{\Psi}_\inc(x)\ga\Psi_\inc(x)\!:$ one finds
\begin{equation}
 j_\inc(s,l)=e\int\frac{v}{v\cdot l}\,\rho_\inc(v)\,d\mu(v)\,,
\end{equation}
so
\begin{equation}
 \int A(x)\cdot J_\inc(x)\,d^4\!x=-\frac{e}{2\pi}\int\frac{v\cdot\Delta V(l)}{v\cdot l}\,d^2\!l\,
 \rho_\inc(v)\,d\mu(v)\,.
\end{equation}
We note that the scaling \eqref{scaling} does not influence this result and the scattering operator \eqref{scop} may be thus written as
\begin{equation}
 S=\exp\left[\,
  -i\int A(x)\cdot J_\inc(x)\,d^4\!x\,\right]\,.
\end{equation}
Therefore the effect of the scaling limit is the omission of time-ordering operator which would normally appear before the exponent.

\section{Discussion}

In the setting of Dirac field scattered by an external classical wave packet we have derived the infrared limit of the scattering operator and confirmed the momentum dependent phase shift first discovered by Staruszkiewicz in a semi-classical approximation. This result has an elevation to the case of quantum Dirac field.

\pagebreak[2]
There are two striking features of these results.\\
 (i) Although a nontrivial asymptotic effect may be observed, the scattering cross-section is trivial (the momentum-representation of the kernel of scattering operator is diagonal in momenta). \\
 (ii) The effect vanishes if the electromagnetic field is infrared-regularized in any way which leads to cutting-off of low frequencies, however small (as then $\tdV(0,l)=0$). This may be read as an indication that also in the full interacting theory infrared regularization may lead to the loss of some of the information. But to avoid such regularization one needs to engage representations of the electromagnetic field which allow for the analogue of non-vanishing characteristic $\tdV(0,l)$. The model of asymptotic fields proposed by the present author, as mentioned in the introduction, may be a candidate for such setting. The work along these lines is in progress.

\end{document}